\documentclass[letterpaper,11pt,twocolumn]{article}
\usepackage{epsfig}
\usepackage{abstract, url}
\usepackage{astjnlabbrev-jh}
\usepackage{amssymb, amsmath, mathtools}
\usepackage{natbib, authblk}
\newcommand\degree{\ensuremath{^\circ}}
\newcommand\degrees\degree

\DeclareSymbolFont{UPM}{U}{eur}{m}{n}
\DeclareMathSymbol{\umu}{0}{UPM}{"16}
\let\oldumu=\umu
\renewcommand\umu{\ifmmode\oldumu\else\math{\oldumu}\fi}
\newcommand\micro{\umu}
\newcommand\micron{\micro m}
\newcommand\microns \micron

\newcommand\arcsec[0]{$^{\prime\prime}$}

\let\oldsim=\sim
\renewcommand\sim{\ifmmode\oldsim\else\math{\oldsim}\fi}
\let\oldpm=\pm
\renewcommand\pm{\ifmmode\oldpm\else\math{\oldpm}\fi}
\newcommand\by{\ifmmode\times\else\math{\times}\fi}

\newbox{\wdbox}
\renewcommand\c{\setbox\wdbox=\hbox{,}\hspace{\wd\wdbox}}
\renewcommand\i{\setbox\wdbox=\hbox{i}\hspace{\wd\wdbox}}

\newcount\timect
\newcount\hourct
\newcount\minct
\newcommand\now{\timect=\time \divide\timect by 60
         \hourct=\timect \multiply\hourct by 60
         \minct=\time \advance\minct by -\hourct
         \number\timect:\ifnum \minct < 10 0\fi\number\minct}

\catcode`@=11

\newcommand\comment[1]{}
\newcommand\commenton{\catcode`\%=14}
\newcommand\commentoff{\catcode`\%=12}

\renewcommand\math[1]{$#1$}
\newcommand\mathshifton{\catcode`\$=3}
\newcommand\mathshiftoff{\catcode`\$=12}

\comment{the backslash is necessary}

\comment{alignment tab}

\let\atab=&
\newcommand\atabon{\catcode`\&=4}
\newcommand\ataboff{\catcode`\&=12}

\let\oldmsp=\sp
\let\oldmsb=\sb
\def\sp#1{\ifmmode
           \oldmsp{#1}%
         \else\strut\raise.85ex\hbox{\scriptsize #1}\fi}
\def\sb#1{\ifmmode
           \oldmsb{#1}%
         \else\strut\raise-.54ex\hbox{\scriptsize #1}\fi}
\newbox\@sp
\newbox\@sb
\def\sbp#1#2{\ifmmode%
           \oldmsb{#1}\oldmsp{#2}%
         \else
           \setbox\@sb=\hbox{\sb{#1}}%
           \setbox\@sp=\hbox{\sp{#2}}%
           \rlap{\copy\@sb}\copy\@sp
           \ifdim \wd\@sb >\wd\@sp
             \hskip -\wd\@sp \hskip \wd\@sb
           \fi
        \fi}
\def\msp#1{\ifmmode
           \oldmsp{#1}
         \else \math{\oldmsp{#1}}\fi}
\def\msb#1{\ifmmode
           \oldmsb{#1}
         \else \math{\oldmsb{#1}}\fi}
\def\supon{\catcode`\^=7}
\def\supoff{\catcode`\^=12}
\def\subon{\catcode`\_=8}
\def\suboff{\catcode`\_=12}
\def\supsubon{\supon \subon}
\def\supsuboff{\supoff \suboff}

\newcommand\actcharon{\catcode`\~=13}
\newcommand\actcharoff{\catcode`\~=12}

\newcommand\paramon{\catcode`\#=6}
\newcommand\paramoff{\catcode`\#=12}

\comment{And now to turn us totally on and off...}

\newcommand\reservedcharson{\commenton \mathshifton \atabon \supsubon \actcharon
	\paramon}

\newcommand\reservedcharsoff{\commentoff \mathshiftoff \ataboff
	\supsuboff \actcharoff \paramoff}

\catcode`@=12
\reservedcharsoff

\reservedcharson

\comment{ Must have ONLY ONE of these... trust these macros, they work

}

\newenvironment{packed_item}{
\begin{itemize}
   \setlength{\itemsep}{1pt}
   \setlength{\parskip}{0pt}
   \setlength{\parsep}{0pt}
}{\end{itemize}}

\newcommand{\squishlist}{
 \begin{list}{$\bullet$}
  { \setlength{\itemsep}{1pt}
     \setlength{\parsep}{0pt}
     \setlength{\topsep}{3pt}
     \setlength{\partopsep}{0pt}
     \setlength{\leftmargin}{2.0em}
     \setlength{\labelwidth}{1.5em}
     \setlength{\labelsep}{0.5em} } }

\newcommand{\squishend}{
  \end{list}  }

\newcommand\arcdeg{\ensuremath{^\circ}}
\newcommand\ie{i.\@e.}
\newcommand\etc{etc.\@}
\newcommand\au{AU}

\reservedcharson
\actcharon

\comment{
\makeatletter
\renewcommand{\@make@caption@text}[2]{%
  \begin{center}
    \makebox[\linewidth]{\rmfamily#1\quad#2}
  \end{center}
}%
\makeatother
}

\title{NEOKepler: Discovering Near-Earth Objects Using the Kepler Spacecraft}

\author[1,4]{Kevin B. Stevenson}
\author[1]{Daniel Fabrycky}
\author[2]{Robert Jedicke}
\author[3]{William Bottke}
\author[2]{Larry Denneau}
\affil[1]{\small Dept.~of Astronomy and Astrophysics, University of Chicago, 5640 S Ellis Ave, Chicago, IL 60637, USA}
\affil[2]{\small Institute for Astronomy, University of Hawaii, 2680 Woodlawn Drive, Honolulu, HI 96822, USA}
\affil[3]{\small Southwest Research Institute, 1050 Walnut Street, Suite 300, Boulder, CO 80302, USA}
\affil[4]{\small kbs@uchicago.edu}

\begin{document}

\twocolumn[
\maketitle

\begin{onecolabstract}

The Kepler Mission has been an irrefutable success.  In the last 4.5 years, it has monitored 150 confirmed exoplanets in over 75 stellar systems and detected an additional $\sim$3,300 planet candidates.  Using these data, we have learned the size distribution of planets in our galaxy, the likelihood that a star hosts an Earth-sized planet, and the percentage of stars that contain multi-planet systems.  The recent failure of a second reaction wheel has ended {\em Kepler's} primary mission; however, its plight is a unique opportunity to make significant advances in another important field, without the time and costs associated with designing, building, and launching another spacecraft.

We propose a new Kepler mission, called NEOKepler, that would survey near Earth's orbit to identify potentially hazardous objects (PHOs).  To understand its surveying power, {\em Kepler's} large field of view produces an {\em etendue} ($A\Omega$) that is 4.5 times larger than the best survey telescope currently in operation.  In this paper, we investigate the feasibility of NEOKepler using a double ``fence post'' survey pattern that efficiently detects PHOs.  In a simulated 12-month survey, we estimate that NEOKepler would detect $\sim$150 new NEOs with absolute magnitudes of less than 21.5, $\sim$50 of which would be new PHOs.  This would increase the annual PHO discovery rate by at least 50\% and improve upon our goal of discovering 90\% of PHOs by the end of 2020.  Due to its heliocentric orbit, {\em Kepler} would also be sensitive to objects inside Earth's orbit, discovering more objects in its first year than are currently known to exist.  Understanding this undersampled sub-population of NEOs will reveal new insights into the actual PHO distribution by further constraining current NEO models.  As an alternative science goal, NEOKepler could employ a different observing strategy to discover suitable targets for NASA's Asteroid Redirect Mission. 
\\

\end{onecolabstract}
]

\section{Science Justification}

Although {\em Kepler's} remarkably successful mission to detect exoplanets has now come to an end, a new and exciting mission awaits. Here we advocate that {\em Kepler} be given the task of finding near-Earth objects (NEOs). NEOs are of great interest to NASA, partly because of the hazard they represent to human life/property, but also because of their scientific potential. By understanding the NEO population, and combining these data with our knowledge of NEO source regions and NEO dynamics, we can glean insights into the delivery processes of meteorites as well as the evolution of asteroids, the surviving building blocks of the terrestrial planets.

All of the attention devoted to these bodies has led to a golden age for NEO research. Over the last two decades, we have discovered 95\% of the most threatening NEOs \citep[those larger than 1~km in diameter,][]{Mainzer2012}, while ongoing surveys (e.g., Catalina Sky Survey) are finding many sub-km NEOs as well. NEO physical characterization studies by missions (e.g., NEAR-Shoemaker), space-based telescopes (e.g., {\em WISE} and {\em Spitzer}), and ground-based observatories (e.g., Arecibo), are also revolutionizing our ideas about what NEOs are actually like.  On 15 April 2010, President Obama announced that NASA would send astronauts to an NEO by 2025; this remains Administration policy and scientists are currently working towards this goal. The 15 February 2013 explosion of a previously-undetected meteor over Chelyabinsk, Russia, has further boosted interest in NEOs.

In our white paper, we devise a preliminary observing strategy and apply numerical models to assess {\em Kepler's} ability to detect NEOs.  Our work indicates that NEOKepler would add considerably to our existing knowledge of NEOs, and could potentially be a ``game changer'' in terms of two particular NEO sub-populations: potentially-hazardous objects (PHOs) and interior-Earth objects (IEOs).

A PHO is an NEO with the potential to make close approaches to the Earth and sufficient mass to cause significant damage in the event of an impact.  It has a minimum orbit intersection distance (MOID) of $<0.05$ AU and its diameter is $>140$ meters.  Using a survey pattern optimized for the cadence of PHOs, particularly those most likely to strike the Earth in the near future, we find that NEOKepler would be about twice as efficient at finding PHOs than the best ground-based observatory.  By itself, {\em Kepler} would singlehandedly increase the annual PHO discovery rate by 50\%.  This will help NASA fulfill the George E. Brown, Jr. Near-Earth Object Survey Act, which directs NASA to discover, track, catalog, and characterize the physical characteristics of at least 90\% of potentially hazardous NEOs larger than 140 meters in diameter by the end of 2020. As of 2012, an estimated 20 to 30\% of these objects have been found.

IEOs are a curious and poorly understood sub-component of the NEO population that are found inside Earth's orbit. Like NEOs, these bodies are derived from the main asteroid belt, but many have followed tortuous dynamical pathways to escape the gravitational clutches of the Earth. These bodies were likely subject to physical processes that have caused mass shedding or even disruption (e.g., tidal disruption via a close pass with Earth/Venus; mass shedding via thermal effects from the Sun or the so-called YORP non-gravitational spin up forces). Our work shows that {\em Kepler} would find as many IEOs in one year as are currently known (i.e., it will double the known population in its first year).

Using a search pattern and cadence optimized for objects on nearly-circular Earth-like-orbits, {\em Kepler} could also be used to search for targets suitable for NASA's ``Asteroid Redirect Mission'' (ARM), in which a robotic spacecraft would intercept and capture a 5-10~m diameter NEO and return it to an orbit within the Earth-Moon system. While this could lead to the development of a potential target list for both robotic and human missions, doing so would prevent {\em Kepler} from focusing on the NEO populations discussed above.

\section{Project Goals}

The goals of the proposed NEOKepler project are as follows:
\begin{packed_item}
\item To discover NEOs that may be classified as being potentially hazardous,
\item To substantially increase the known IEO population,
\item To test theoretical predictions of NEO distributions,
\item To search for suitable targets for the Asteroid Redirect Mission (ARM), and
\item To work with NEOWISE and ground-based programs to further constrain the sizes and orbits of new candidate detections and previously-known NEOs.
\end{packed_item}
Employing the observing strategy described in Section \ref{sec:projdesc} (Method 2) in a simulation, we anticipate that a 12-month mission will successfully detect $\sim$860 NEOs with absolute magnitudes $H<21.5$.  One quarter of these objects would have a MOID of $<0.05$ AU, classifying them as PHOs, and 50 of those would be new discoveries.  The simulations also predict a twofold increase in the number of known IEOs, which have aphelion distance $Q<0.983$~AU.  These objects are extremely challenging to detect from the Earth or an Earth-bound orbit.  With a statistically significant number of identified IEOs, we can compare the unbiased NEO distribution to theoretical predictions \citep{Bottke2002}.  We can also perform a similar comparison with the undersampled population of high-inclination NEOs.

Although {\em Kepler} is restricted from pointing towards opposition, where NEOs appear brightest, it may still be able to identify suitable targets for the ARM.  A potentially detectable 10-meter-class object would need to pass within 0.02~AU of the spacecraft.  Unfortunately, nearby objects have large apparent velocities and fade relatively quickly, thereby making it difficult to precisely determine their orbits using only a few exposures.  {\bf With additional research and simulations, observing strategies specifically geared towards identifying potential ARM targets may be employed to increase their probability of detection.}

In an extended mission (called NEOWISE) from October 2010 to February 2011, NASA's Wide-field Infrared Survey Explorer (WISE) searched for and found 134 NEOs.  NASA recently announced that in September of 2013, NEOWISE will commence operations for an additional three years with the goal of discovering 150 new NEOs and characterizing 2,000 known objects.  NEOKepler, which is sensitive to reflected light from NEOs, could work with NEOWISE, which is sensitive to their thermal emission, to constrain the albedos of objects along a similar line of sight.  Currently, NEO albedos are poorly constrained and add a significant uncertainty to object size estimates.

One method of measuring the power of a survey telescope is called the {\em etendue}, which is defined as the collecting area, $A$, multiplied by the field of view, $\Omega$.  We list the etendue for several existing survey instruments in Table \ref{tab:etendu}.  {\bf {\em Kepler} has, by far, the largest etendue of all existing surveys.}  Additionally, unlike ground-based surveys, NEOKepler would be capable of making continuous observations without interruption from daylight or bad weather.  This is particularly important when tracking objects over days and sometimes weeks.

\begin{table}[htb]
\centering
\caption{\label{tab:etendu}
Survey Telescope Power.}
\begin{tabular}{@{}l@{}ccc@{}}
    \hline
    \hline
    Telescope       & Diameter  & FOV               & $A\Omega$ \\
                    & [m]       & [\degrees\sp{2}]  & [(m\degrees)\sp{2}]  \\
    \hline
    WISE            & 0.4       & 0.6               &  0.08     \\
    CFHT Megacam    & 3.6       & 1.0               & 10.2      \\
    SDSS            & 2.5       & 3.0               & 14.7      \\
    Pan-STARRS (PS1)& 1.8       & 7.0               & 17.8      \\
    Kepler          & 0.95      & 115               & 81.5      \\
    \hline
\end{tabular}
\end{table}

\section{Project Description}
\label{sec:projdesc}

In the subsections below, we describe the details of our proposed project.  In overview, NEOKepler would implement a double ``fence post'' survey pattern, with each post being 20{\degrees} in longitude and 120{\degrees} in latitude.  The spacecraft would acquire 72 full frame images (FFIs) per survey region, alternating three times between pairs of {\em Kepler} fields to detect NEO motion on the scale of $\sim$90 minutes, thus forming a tracklet.  {\em Kepler} would return to each survey region every 4.5 days.  We estimate the mean rate of NEO motion within the field to be 0.74\degrees/day, allowing {\em Kepler} to acquire six tracklets before a typical NEO passes through a survey region.  To avoid transmitting FFIs over the Deep Space Network (DSN), an on-board trail-finding algorithm would identify candidate detections, trim and save those subregions to the solid state recorder (SSR), and discard the remaining FFI.  Ground-based resources would then be used to link tracklets and derive NEO objects.

\subsection{Target Field}

\begin{figure}[tb]
\includegraphics[width=1.0\linewidth]{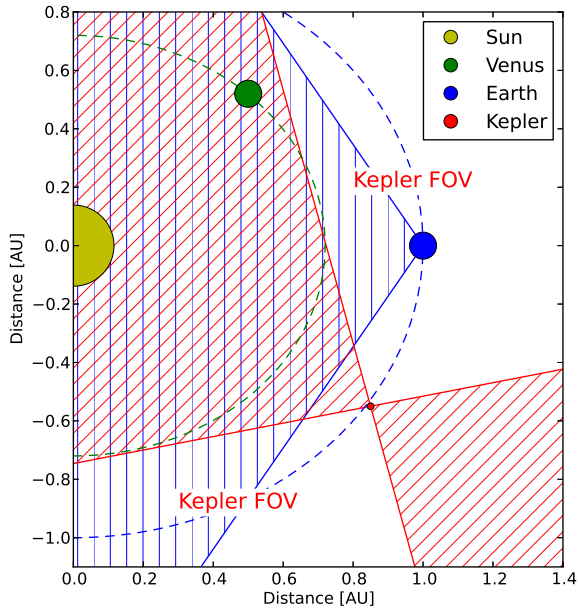}
\caption{\label{fig:fov}\small
\textbf{{\em Kepler's} position and field of view (FOV) within the ecliptic plane.} The blue and red hashed regions represent areas of the ecliptic that are inaccessible to NEO survey missions from Earth and {\em Kepler}, respectively.  Due to its separation from Earth, {\em Kepler} is capable of discovering IEOs and PHOs not readily detectable from Earth.
}
\end{figure}

{\em Kepler} is on an Earth-trailing heliocentric orbit with a period of 372.5 days.  This causes the spacecraft to slowly drift away, as seen from Earth.  After 4.5 years, {\em Kepler} is now $\sim0.5$~AU from Earth, forming a $\sim30\degree$ angle with respect to the Sun.  Figure \ref{fig:fov} depicts a representative view of their positions within the ecliptic plane.  

As discussed in the Call for White Papers, {\em Kepler's} pointing is restricted to 90 {\pm} 45{\degrees} away from the Sun.  Thus, the spacecraft is forbidden to search for NEOs at or near opposition.  However, {\em Kepler's} position gives it a significant advantage over Earth-bound NEO surveys: {\em Kepler} is sensitive to IEOs currently located between the Earth and the Sun (see Figure \ref{fig:fov}).  These objects would otherwise be undetectable during daylight hours from the Earth.

For this project, we propose two survey regions for detecting NEOs.  The primary region would be directed towards the Earth near $+70\arcdeg$ solar elongation and would be composed of a 2$\times$12 grid of {\em Kepler} fields encompassing 20{\degrees} in longitude and 120{\degrees} in latitude.  The secondary survey region would be directed back along Earth's orbit near $-70\arcdeg$ solar elongation and would have the same dimensions.  Due to the detectors' unique configuration, the Earth-Moon system can easily be placed at any corner within {\em Kepler's} field of view without risk of saturating the detectors.

\subsection{Pointing Stability}

We estimate {\em Kepler's} pointing stability for this project using the information contained in the Exoplanetary Appendix to the Kepler Project call for white papers \citep{KeplerAppendix2013}.  The spacecraft achieves maximum pointing stability when the boresight is pointed in the velocity- or anti-velocity-vectors.  At these pointings, boresight drift is $<0.1${\arcsec} ($<0.025$ pixels) for a 12-minute observation, assuming a 0.005{\degree} initial offset error.  Pointings perpendicular to the orbital plane can achieve similarly manageable drift rates over the same frame time.  Our two proposed survey regions are centered only 20{\degrees} from the optimal pointing-stability plane.  Within these regions, continuously repointing the telescope after each frame does not allow sufficient time for torque to accumulate to a secular drift; therefore, we conclude that boresight drift will not adversely affect our science goals.

\subsection{Data Acquisition, On-Board Processing, and Transmission}

There are a multitude of techniques one might consider employing to efficiently acquire, process, and transmit the data back to Earth.  Here we present two methods: with and without the need for software modifications.  These techniques can be used as a spring board in future studies to determine their efficacy given {\em Kepler's} constraints and the level of funding available to make these modifications.

\subsubsection*{Method 1}

A simple version of this project may be accomplished with the current flight software.  At each field, {\em Kepler} would acquire a 12-minute FFI.  While the FFI is writing to the SSR, which takes 20 minutes, the telescope would slew to an adjacent field and acquire another 12-minute FFI.  {\em Kepler} would observe this pair of fields three times in an ABABAB pattern before moving to the next pair.  Within each set of three images, pairs of NEO detections would be separated in time by $\sim$90 minutes.  
The SSR can hold 42 FFIs total before the on-board memory is full, requiring a data download after observing only seven different pairs of fields (every $\sim$32 hours).   All of the processing required for asteroid detection and tracklet determination would be done on the ground.  The price for software simplicity is both efficiency and DSN time: the downloads of the data would take more than 50\% of the total time.

\subsubsection*{Method 2}

A more sophisticated, and preferable, survey would include software elements that allow for on-board processing.  Below, we describe which hardware and algorithms exist to carry out each element.

A simple yet efficient observing strategy would be to employ asymmetric windowing.  Here, a 12-minute exposure is subdivided into 2 minutes of acquisition, a 5-minute break, and then an additional 5 minutes of acquisition.  We estimate the motion of a typical NEO to be $<1$ pixel every 100 seconds.  Thus, a 5-minute break would produce at least a 3-pixel gap in the trail.  The advantage of this technique is that a single FFI would still give directional information about the NEO trail.  The hardware to accomplish this is already in place, through differencing of the tally of the science data accumulator which generates the short cadence data.  The same technique can be used to omit some of the integrations to create the asymmetric gap.  As with Method 1, {\em Kepler} would observe each pair of fields three times and the telescope would slew between pairings while the FFI is being written to the SSR.

On board {\em Kepler}, there is a CPU that requantizes and compresses the data before writing to the SSR.  This time and equipment could instead be used for two additional software elements.

First, the software would perform image subtraction between each frame and a pseudo-master frame for that field.  Storing 48 master frames requires more on-board SSR memory than is available.  A pseudo-master frame could be constructed using the median of the three recently-acquired FFIs for each field.  Subtracting a master frame may not fully remove bright sources -- trails that overlap such sources may not be detected anyway.  However, subtraction does reduce the influence of the multitude of faint sources that exist over a significant fraction of the survey area.  With image subtraction, if a 20\sp{th} magnitude NEO trail lies on a 19\sp{th} magnitude galaxy, we would still be able to detect the NEO.

The second, more intensive software element is a trail-finding algorithm.  There is an extensive literature within the computer vision community on Hough transforms, which can detect such trails in a 2-dimensional image (or trails in a stack of such images, a 3-dimensional analogy).   Given that the trail length is a free parameter, the algorithm can efficiently group operations (adding the signal of short trails to make long trails) to run in $\mathcal{O}(N\log N)$ time \citep{2005Arias}.  The results of such algorithms have already been employed for asteroid detection \citep{2007Kubica}.  One of the first steps in assessing the feasibility of the software-intensive portion of this proposal should be to study how long these algorithms would take to execute on board {\em Kepler}.

The result of this sequence of operations would be a time-tagged set of pixels containing candidate NEO trails (known as detections), with astrometric annotation (resulting from the star alignment step for image subtraction).  These data would constitute a very small fraction of the full frame image, thus the mission could conduct several hundred pointings before transmitting the data back to Earth.  These software elements may involve a modest overhead of computation time, but would economise the DSN time.  Once on the ground, the candidate detections would be run through linking software on more powerful computers to identify their orbits and confirm their reality.

\subsection{Ground-Based Data Reduction and Analysis}

Ground-based processing of NEOKepler detections would follow procedures described for the Pan-STARRS Moving Object Processing System \citep[MOPS,][]{Denneau2013}. MOPS is an existing, proven software package that can identify asteroids from raw detection streams. Under the MOPS processing model for NEOKepler, detections would be ingested into MOPS, grouped into tracklets, and stored in the MOPS relational database.  MOPS associates tracklets over 4.5-day intervals into linkages called `derived objects' from which asteroid orbits and absolute magnitudes are computed. NEOKepler derived object parameters and their constituent raw detections would be delivered electronically to the IAU Minor Planet Center for inclusion in the MPC's small body database and for further hazard analysis. The determination of whether a NEOKepler derived object is a new discovery is performed by the MPC, so no filtering of known objects by NEOKepler or its MOPS processor is necessary.

The hardware requirements for MOPS are modest; a small cluster consisting of a master control system, several processing nodes, and a database server can easily handle the data volume from NEOKepler.

\section{Expected Results}

\subsection{Detection, Tracking, and Size Limits}

\begin{figure}[tb]
\includegraphics[width=1.0\linewidth]{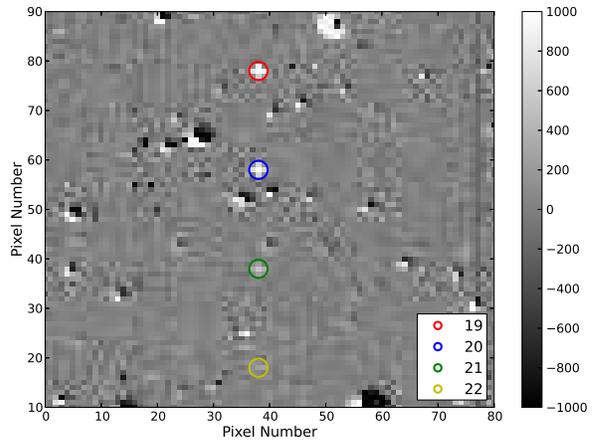}
\caption{\label{fig:ffidetect}\small
\textbf{Detecting NEOs in a differenced FFI.}  We insert four simulated NEOs into one frame with apparent magnitudes 19 (red), 20 (blue), 21 (green), and 22 (yellow).  The total exposure time is 40 seconds per FFI.
}
\end{figure}

We estimate the NEOKepler detection limiting magnitude using publicly-available FFIs from 7 February and 6 March 2013.  We apply a 40-second exposure time to small subregions from each frame before inserting into one of the subregions four simulated NEOs with apparent magnitudes, $m$, from 19 to 22.  During this time, a typical NEO will have moved $<0.5$ pixels on the detector.  To achieve the optimal subtraction between the two subregions, we interpolate each image to high resolution, shift one over the other, and locate the position that minimizes the sum of the absolute difference.  Figure \ref{fig:ffidetect} displays the resulting differenced image and Table \ref{tab:snr1} gives the expected S/N per 40-second exposure.  With this frame time, we estimate that NEOKepler can detect objects as faint as $m\sim22$.

\begin{table}[tb]
\centering
\caption{\label{tab:snr1}
NEO Detection Signal-to-Noise Ratios.}
\begin{tabular}{ccccc}
    \hline
    \hline
    $m$     & Flux      & $\sigma$\sb{BG}   & $\sigma$\sb{Poisson}  & S/N   \\
            & [e\sp{-}] \\
    \hline
    19      & 12,583    & 147               & 112                   & 68    \\
    20      &  5,010    & 147               &  72                   & 31    \\
    21      &  1,994    & 147               &  45                   & 13    \\
    22      &    794    & 147               &  28                   &  5.3  \\
    \hline
\end{tabular}
\end{table}

Estimating tracking limits is complicated by trailing losses, where light from an object is spread along the length of its trail and each pixel receives less light than it would have had the object remained stationary.  Furthermore, once light from a moving object no longer contributes to a given pixel, that pixel's S/N begins to decrease as the background flux continues to contribute to the uncertainty.  Thus, we achieve our highest S/N detections using exposure times on the order of an object's apparent rate of motion on the detector.

\begin{figure}[tb]
\includegraphics[width=1.0\linewidth]{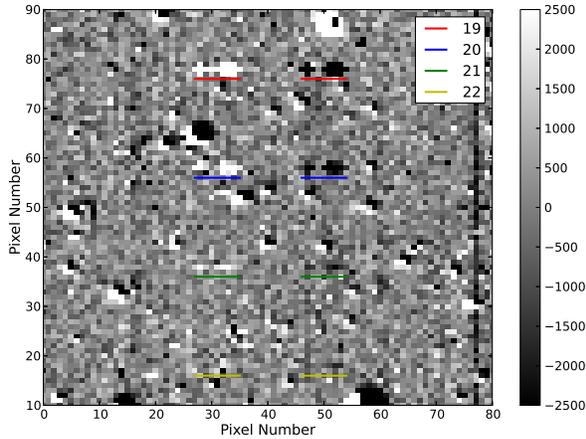}
\caption{\label{fig:ffisim}\small
\textbf{Tracking NEOs in a differenced FFI.}  We insert four simulated NEOs moving from left to right with apparent magnitudes 19 (red), 20 (blue), 21 (green), and 22 (yellow).  NEOs appear lighter (darker) at positions within the first (second) FFI.  The small gaps in the NEO trails are due to asymmetric windowing; the large gaps represent the time needed for data transfer between FFI acquisitions.  The effective exposure time is 7 minutes per FFI.
}
\end{figure}

To estimate the NEOKepler tracking limits, we apply a 7-minute effective exposure time to the same FFI subregions discussed above.  We assume that the NEOs move at a rate of $\sim1$\degree/day and insert a 20-minute gap between FFIs to simulate the time needed for data transfer to the SSR.  Figure \ref{fig:ffisim} displays the differenced image and Table \ref{tab:snr} gives the expected S/N per differenced image.  Using the observing strategy described in Section \ref{sec:projdesc} (Method 2), we estimate that NEOKepler can reliably track objects with $m\sim21$, and possibly fainter.

\begin{table}[tb]
\centering
\caption{\label{tab:snr}
NEO Tracking Signal-to-Noise Ratios.}
\begin{tabular}{ccccc}
    \hline
    \hline
    $m$     & Flux      & $\sigma$\sb{BG}   & $\sigma$\sb{Poisson}  & S/N   \\
            & [e\sp{-}] \\
    \hline
    19      & 251,657   & 660               & 502                   & 67    \\
    20      & 100,186   & 660               & 317                   & 27    \\
    21      &  39,885   & 660               & 200                   & 11    \\
    22      &  15,878   & 660               & 126                   &  4.3  \\
    \hline
\end{tabular}
\end{table}

Assuming a tracking magnitude limit of $m\sim21$, in Table \ref{tab:sizes} we estimate the absolute magnitude limit at various separations assuming both objects are at 1~AU from the Sun.  Then, using a typical set of albedo values, we estimate the limiting range of possible object sizes.  {\bf At the current Earth-{\em Kepler} separation distance of 0.5~AU, NEOKepler could track PHOs as small as 150~m.  Additionally, {\em Kepler} could detect potential ARM targets within 0.02~AU of the spacecraft.}

\begin{table}[tb]
\centering
\caption{\label{tab:sizes}
NEO Size Limits.}
\begin{tabular}{ccc}
    \hline
    \hline
    Distance    & $H$       & Diameter      \\
    \@[AU]      &           & [meters]      \\
    \hline
    1.0         & 20.6      & 210 -- 470    \\
    0.5         & 21.2      & 150 -- 350    \\
    0.2         & 22.9      &  65 -- 150    \\
    0.1         & 24.3      &  35 --  80    \\
    0.05        & 25.8      &  18 --  38    \\
    0.02        & 27.8      &   7 --  15    \\
    0.01        & 29.3      &   3 --   8    \\
    \hline
\end{tabular}
\end{table}

\subsection{PHO Survey Simulation}
\label{sec:survey}

\begin{figure}[tb]
\includegraphics[width=1.0\linewidth]{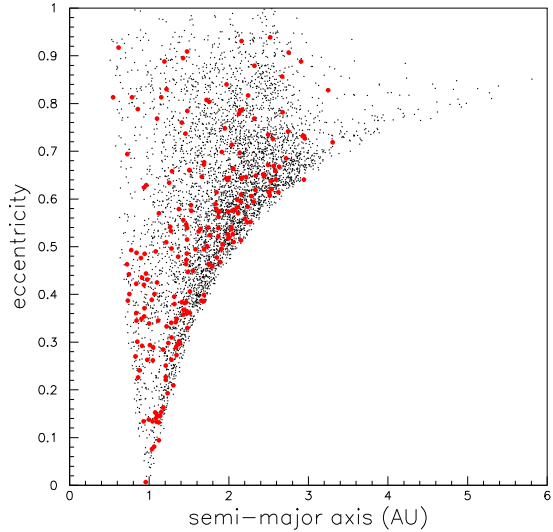}
\caption{\label{fig:kepler-pho-ae-dist}\small
\textbf{Eccentricity vs.~semi-major axis distribution for PHOs discovered in our Kepler simulation (red) and the model PHO population (black).}  The right edge of the distribution is roughly bound by objects with perihelion at 1.05~AU while the left edge represents objects with aphelion at 0.95~AU.
}
\end{figure}

\begin{figure}[tb]
\includegraphics[width=1.0\linewidth]{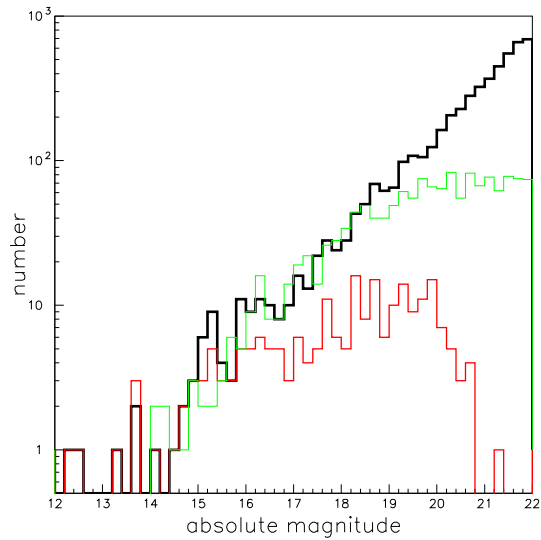}
\caption{\label{fig:kepler-pho-h-dist}\small
\textbf{Absolute magnitudes of PHOs in the model (black), known population (green) and detected in the 1 year synthetic Kepler simulation (red).}  The good agreement between the model and the known population for $H\lesssim18$ is a testament to the fidelity of the model and shows that most PHOs with $H<18$ are already known.  Kepler's contribution to the PHO discovery effort is in the range $H\gtrsim18.5$.
}
\end{figure}

We performed a simulation of a 1-year Kepler PHO survey using the best available model population from \citet{Bottke2002} and the Moving Object Processing System (MOPS) of \citet{Denneau2013}.
The model contained 3,236 PHOs with realistic semi-major axis, eccentricity, inclination, and absolute magnitude distributions (see Figures~\ref{fig:kepler-pho-ae-dist} and \ref{fig:kepler-pho-h-dist}).  The angular orbital elements were assumed to be randomly and evenly distributed in the range $[0\arcdeg,360\arcdeg)$.  

\begin{figure*}[tb]
\centering
\includegraphics[width=0.8\linewidth]{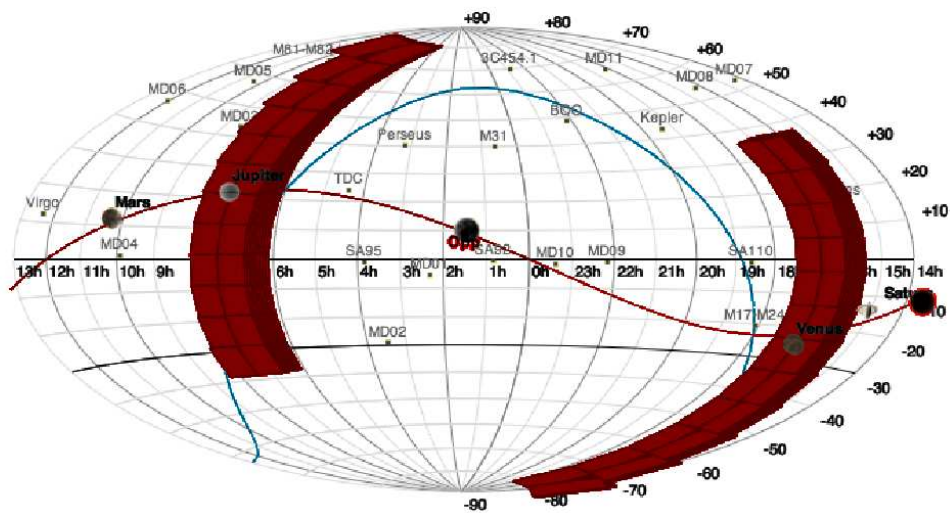}
\caption{\label{fig:kepler-survey-pattern}\small
\textbf{NEOKepler survey pattern.} The highlighted red squares show all of the fields acquired in one lunation of the survey.
}
\end{figure*}

The simulation implemented the double ``fence post'' survey pattern described in Section \ref{sec:projdesc}, which is designed to enhance the detection of PHOs based on our experience with asteroid surveys and the work of \citet{Chesley2004}.  As illustrated in Figure~\ref{fig:kepler-survey-pattern}, each post is a $2\times10$ Kepler field stack of images.  The posts were centered at $\pm70\arcdeg$ solar elongation such that Earth would lie at the center of one post, as viewed from {\em Kepler}.  Mimicking the proposed survey pattern, the simulation visited pairs of adjacent fields three times each until all of the fields in the pattern were visited.  The pattern repeats every 4.5~days, ensuring that each PHO remains in the survey region over many cycles.
  
PHOs should be enhanced relative to the overall NEO population in this survey because it preferentially images the volume within 0.05~AU of Earth's orbit.  We extended the pattern far to the north and south to avoid the ground-based survey bias of preferentially surveying along the ecliptic.  Furthermore, the survey should be sensitive to IEOs because a substantial portion of the search volume is interior to Earth's orbit.

The limiting magnitude was set to $m=22.0$ (corresponding to $S/N=4.3$) with 100\% efficiency to that value and 0\% efficiency for fainter objects.  MOPS calculates the precise location of each PHO in each field, fuzzes the astrometry and photometry, then links detections from the three visits to each field into `tracklets.'  MOPS then links detections of the same object's tracklets acquired on different repeats of the survey pattern into `tracks' that are then fit to a heliocentric orbit.  If the orbit fit is good then the track becomes a `derived object' with an orbit and absolute magnitude.  On subsequent nights, even after half a year, MOPS attempts to `attribute' new tracklets to known derived objects to extend the `length' of the orbit arc (the arc length is typically measured in days).  Furthermore, when a new derived object is created or its orbit is updated because of an attribution, MOPS attempts to `precover' tracklets in the database that have not already been associated with known derived objects.  Since MOPS was created for ground-based surveys and {\em Kepler} is on an Earth-orbit-like trajectory, this MOPS simulation was performed from the location of the Pan-STARRS PS1 telescope on Haleakala, HI, with little loss in fidelity for the purpose of this study.

The efficiency of every step in this simulation was very high, ranging from $\sim100$\% for tracklet creation to 95\% for attribution and precovery.  These results are consistent with our experience and we think they are applicable to a real operating system for which the false detection rate is $\lesssim100\times$ the actual detection rate and for which the systematic false detection rate is relatively small (\ie\ detections due to systematic field artifacts such as edge effects, diffraction spikes, \etc).

\begin{figure}[tb]
\includegraphics[width=0.9\linewidth, angle=90]{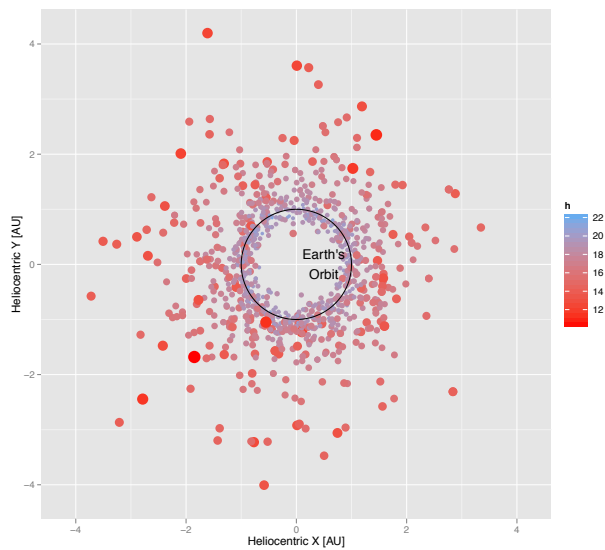}
\caption{\label{fig:kepler-neo-posns}\small
\textbf{Positions of {\em Kepler} NEOs at time of discovery.}  The circle colors and sizes depict the NEO absolute magnitudes.
}
\end{figure}

Figure~\ref{fig:kepler-neo-posns} illustrates how the implemented NEOKepler survey excels at detecting NEOs near Earth's orbit.  The survey identifies objects both inside and outside of 1~AU because it is centered at $70\arcdeg$ solar elongation, but the relative ratio can be tuned by modifying the survey pattern.  Since an object's apparent brightness is determined by its size, and heliocentric and {\em Kepler}-centric distances, it is difficult to speculate how changing the survey pattern would affect the number and orbit distribution of the detected objects.  Surveying at larger solar elongations would decrease the phase angle, but would also preferentially detect objects further from the Sun, which are typically fainter.  On the other hand, it is possible that the orbit distribution yields a `richer' NEO sky-plane density at larger solar elongations.  Additional simulations would serve to optimize the survey pattern based on the specified science goals.

The resulting distribution of semi-major axes, eccentricities, and inclinations of the simulated {\em Kepler} PHO discoveries (see Figure~\ref{fig:kepler-pho-ae-dist}) are similar to the ensemble of known PHOs, except that the {\em Kepler} PHOs tend to have:
\begin{packed_item}
\item smaller semi-major axes,
\item higher eccentricities,
\item higher inclinations, and
\item a more model-like PHO distribution, \ie\ suffer {\it less} from observational selection effects, than ground-based surveys.
\end{packed_item}
\noindent As quantified in Table~\ref{tab:pho-percentages}, {\bf the NEOKepler survey is roughly twice as effective as all current surveys at identifying PHOs} with $H<21.5$.  While only $\sim$13\% of known NEOs with $H<21.5$ are PHOs, about 25\% of the NEOs detected with NEOKepler are in the same size range.  In fact, the PHO fraction within the NEOKepler survey is $\sim$20\% larger than the fraction within the actual population.  All of these results are a consequence of our simulated survey pattern, which was chosen to favor PHO detections.  We note that, while higher inclination PHOs have a smaller collision probability with Earth because they spend less time near Earth, their higher relative speed with respect to Earth makes each collision more dangerous since the impact energy increases as the square of the impact speed.

\begin{table}[tb]
\centering
\caption{\label{tab:pho-percentages}
Percentage of NEOs that are PHOs.}
\begin{tabular}{@{}cc@{}}
    \hline
    \hline
    Population                  & PHO \%    \\
    ($H<21.5$, MOID~$<0.05$\au) &           \\
    \hline
    Known NEOs                  &  13\%     \\
    NEOKepler NEOs              &  25\%     \\
    \hline
\end{tabular}
\end{table}

For $H<21.5$, NEOKepler's PHO discovery rate is about twice as high as the best contemporary ground-based survey and would {\bf single-handedly increase the world-wide PHO discovery rate by about 50\%} (see Table~\ref{tab:pho-discovery-rates}).  To estimate the NEOKepler discovery rate, we use the data in Figure~\ref{fig:kepler-pho-h-dist} to measure both NEOKepler's PHO detection efficiency as a function of absolute magnitude, $\epsilon(H)$, and the number of unknown PHOs, $N_{unknown}(H) = N_{model}(H) - N_{known}(H)$.  The annual discovery rate for PHOs with $21.5>H>18.5$ is then $n \sim \int_{18.5}^{21.5} dH \epsilon(H) N_{unknown}(H) \sim 50$.

\begin{table}[tb]
\centering
\caption{\label{tab:pho-discovery-rates}
Annual PHO Discovery Rates.}
\begin{tabular}{@{}c@{}c@{}}
    \hline
    \hline
    System                              & Discovery Rate    \\
    ($H<21.5$, MOID~$<0.05$\au)         &                   \\
    \hline
    All observatories worldwide\sp{1}   &  $\sim90$         \\
    Single best observatory\sp{1}       &  25-30            \\
    NEOKepler survey                    &  $\sim50$         \\
    \hline
\end{tabular}
\begin{minipage}[t]{1.0\linewidth}
\sp{1}{\url{http://www.minorplanetcenter.net/iau/lists/YearlyBreakdown.html}}
\end{minipage}
\end{table}

The {\em Kepler} survey implemented here excels at detecting IEOs relative to ground-based systems, which have difficulty detecting IEOs since they are never more than $90\arcdeg$ in solar elongation (see Table~\ref{tab:ieos}).  Surveying at small solar elongations from the ground is difficult because this area is always close to the horizon, suffers from high-airmass, twilight, bad seeing, obstructions, and is only accessible for a short time after astronomical twilight ends or before it begins.  None of these problems exist in space, so 2\% of the detected objects in our implemented NEOKepler survey are IEOs, compared to just 0.2\% in the predicted NEO population.  {\bf In just its first year of operation, NEOKepler would identify more IEOs than are currently known: $\sim$18 vs.~15.}

\begin{table}[tb]
\centering
\caption{\label{tab:ieos}
12-Month NEOKepler IEO Count.}
\begin{tabular}{cc}
    \hline
    \hline
    Population                  & \#    \\
    ($H<21.5$, $Q<0.983$~AU)    &       \\
    \hline
    Known IEOs\sp{1}            & 15    \\
    NEOKepler IEOs              & 18    \\
    \hline
\end{tabular}
\begin{minipage}[t]{1.0\linewidth}
\sp{1}{\url{http://ssd.jpl.nasa.gov/sbdb_query.cgi}}
\end{minipage}
\end{table}

The NEOKepler survey pattern and simulation described above can be improved in many ways.  A serious trade study would consider issues like total survey area coverage, central solar elongation, longitude and latitude ranges, and the impact each factor has on the realized discovery rate of interesting objects (presumably PHOs and IEOs, though perhaps also candidate ARM targets).  The implemented survey was constructed in R.A. and declination coordinate space so that the solar elongation changes between fields.  A more profitable scenario might be to keep the NEOKepler field centers along lines of constant solar elongation.  In any event, even this simplistic survey illustrates the power that NEOKepler can bring to bear on identifying hazardous and scientifically interesting objects.

\section{Conclusions}

Despite recent setbacks, the successes of the Kepler Mission can continue with the realization of NEOKepler.  This program would survey near Earth's orbit to identify PHOs with a survey telescope power, or etendue, that surpasses all survey telescopes currently in operation.  Using a double ``fence post'' survey pattern, {\bf a 12-month mission would discover $\sim150$ new NEOs with $H<21.5$, including $\sim50$ new PHOs and $\sim18$ new IEOs.}  This would increase the annual PHO discovery rate by at least 50\%, double the number of known IEOs, and improve upon our goal of discovering 90\% of PHOs by the end of 2020.

Furthermore, the current observing strategy could be modified to detect ARM objects, which are currently few in number.  If this goal is determined to be a priority, additional research and simulations would be employed to explore observing strategies that could enhance their detection probability.

{\small
\bibliography{NEOKepler}
}

\end{document}